\documentclass[aps, prl,
 amsmath,
 amssymb,
reprint,
showpacs,
showkeys,
]{revtex4-1}
\usepackage{graphicx}
\pdfoutput=1
\usepackage{dcolumn}
\usepackage{bm}

\begin{document}
\title{Shock instability in dissipative gases}
\author{M. I. Radulescu}
   \email{matei@uottawa.ca}
\author{N. Sirmas}
\affiliation{Department of Mechanical Engineering, University of Ottawa}
\date{\today}

\begin{abstract}
Previous experiments have revealed that shock waves in thermally relaxing gases, such as ionizing, dissociating and vibrationally excited gases, can become unstable.  To date, the mechanism controlling this instability has not been resolved.  Previous accounts of the D'yakov-Kontorovich instability, and Bethe-Zel'dovich-Thompson behaviour could not predict the experimentally observed instability. To address the mechanism controlling the instability, we study the propagation of shock waves in a simple two-dimensional dissipative hard disk molecular model.  To account for the energy relaxation from translational degrees of freedom to higher modes within the shock wave structure, we allow inelastic collisions above an activation threshold.  When the medium allows finite dissipation, we find that the shock waves are unstable and form distinctive high density non-uniformities and convective rolls on their surface.  Using analytical and numerical results for the shock Hugoniot, we show that both DK and BZT instabilities can be ruled out.  Instead, the results suggest that the clustering instability of Goldhirsch and Zanetti in dissipative gases is the dominant mechanism.  
\end{abstract}

\maketitle

\begin{figure}
 \includegraphics[width=.9\columnwidth]{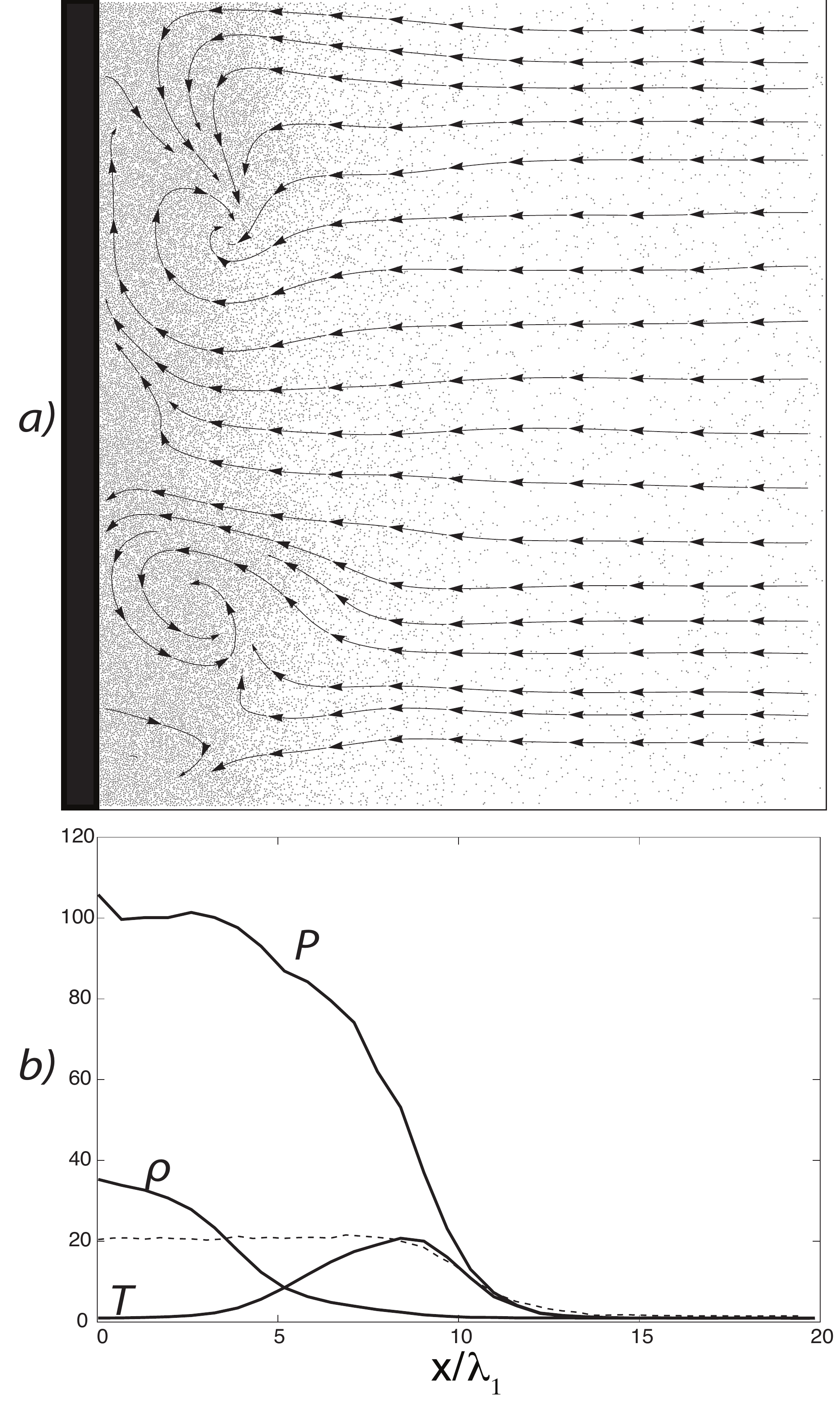}
 \caption{a) Particle distribution and coarse-grained streamlines; b) coarse grained one-dimensional distribution of pressure, temperature and density for $u_p=7$, $\epsilon=0.95$ and $u^*=3$; broken line is the shock temperature distribution for $\epsilon=1$ and the same shock Mach number.}
 \label{fig:stream} 
\end{figure}

\begin{figure}
 \includegraphics[width=.95\columnwidth]{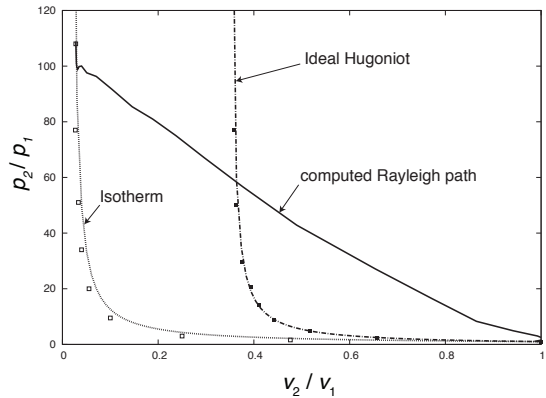}
 \caption{The equilibrium states achieved across a system undergoing elastic (black symbols) and activated inelastic collisions (open symbols) and comparison with the ideal Hugoniot and the isotherm computed with the Helfand equation of state.}
 \label{fig:hugoniot} 
\end{figure}

\begin{figure}
\includegraphics[width=.9\columnwidth]{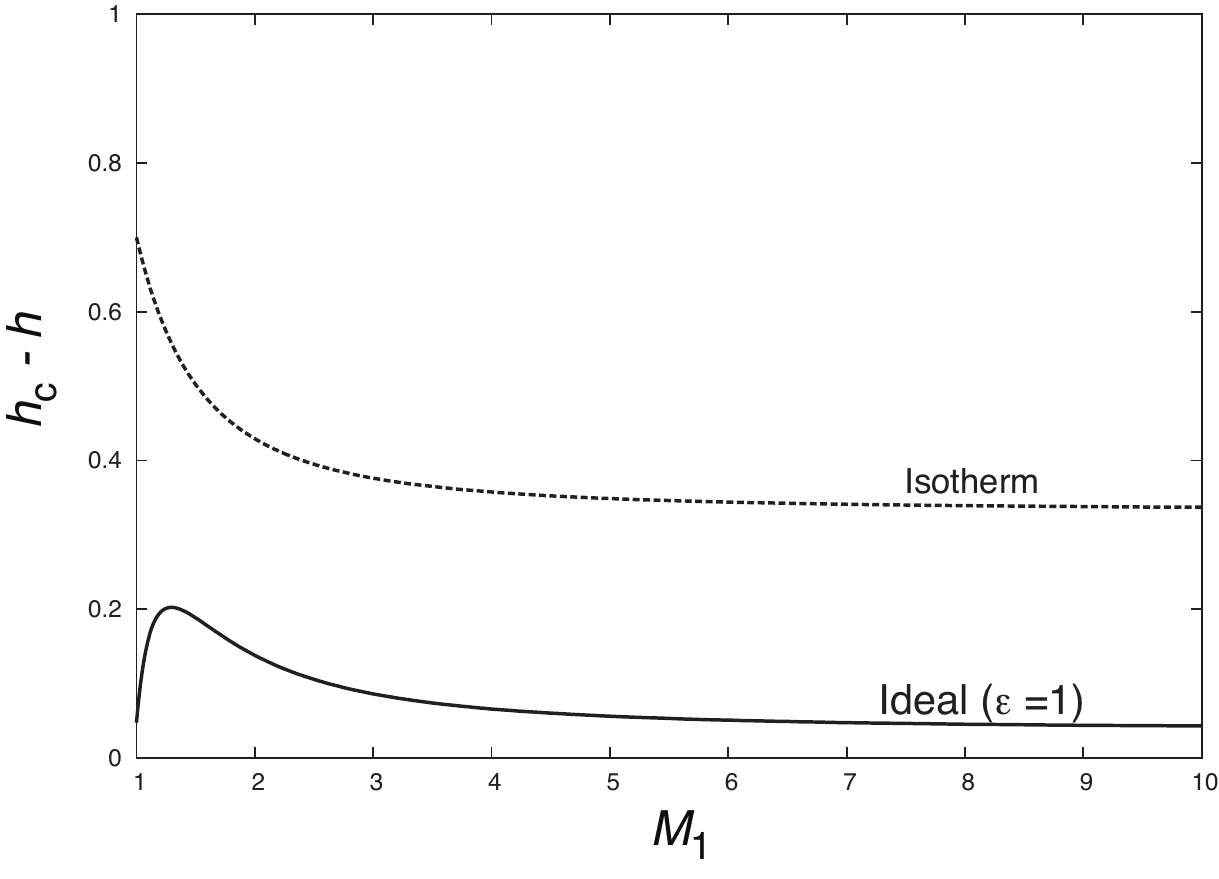}
\caption{Propensity for D'yakov-Kontorovich instability in the ideal and in the  dissipative hard disk gas; $h_c-h < 0$ indicates instability.}
\label{fig:DKcheck} 
\end{figure}

\begin{figure}
\includegraphics[width=.8\columnwidth]{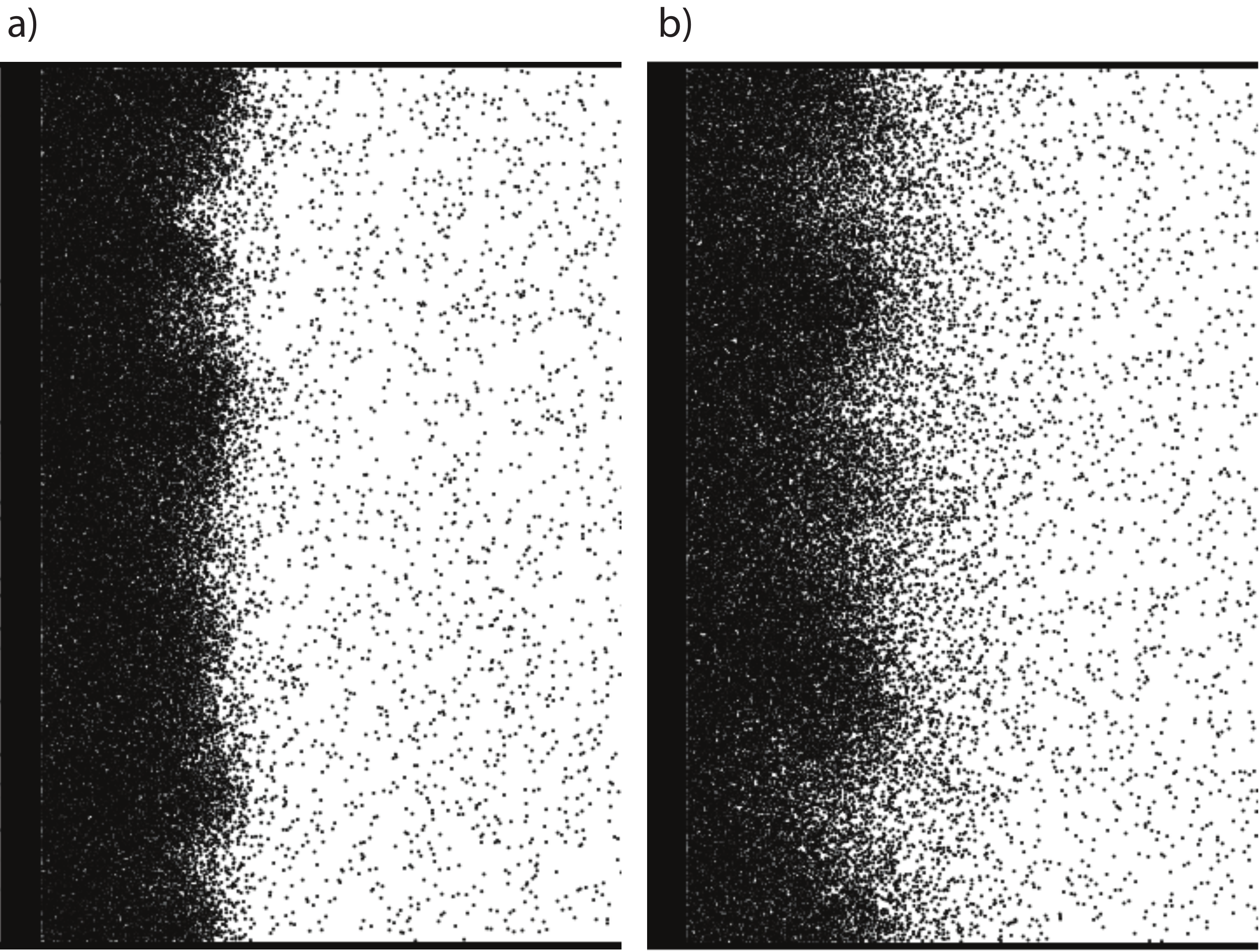}
\caption{Shocked layer morphology in a dissipative gas with a) $\epsilon=0.8$ and b) $\epsilon=0.93$.}
\label{fig:othereps} 
\end{figure}

Shock waves are well known to become unstable when travelling through an \textit{exothermic} medium \cite{Fickett&Davis1979}.  However, locally \textit{endothermic} processes have also been shown to lead to shock instabilities, although their origin and physical mechanism remain unclear.  For example, sufficiently strong shock waves leading to ionization have been shown to develop instabilities \cite{Grunetal1991, Struth1963, Glass&Liu1978, Griffithetal1976}.   Shock instabilities have also been observed in sufficiently strong shock waves when dissociation of the molecules is possible \cite{Griffithetal1976}. Perhaps the most puzzling experimental observations correspond to shock instabilities in carbon dioxide \cite{Griffithetal1976, Mishinetal1981,Hornung&Lemieux2001} and other larger molecules, in which neither ionization nor dissociation is expected.\\
At present, an explanation for these observed instabilities is lacking.  This is the focus of the present Letter. Current known accounts for shock instabilities in non-reactive media are the D'yakov-Kontorovich (DK) instability \cite{Landau&Lifshitz1987}, which leads to undamped emission of sound from the shock and vortex-entropy waves, and the anomalous behaviour in Bethe-Zel'dovich-Thompson fluids \cite{Zeldovich&Raizer1966} which leads to shock splitting.  Griffith et al. \cite{Griffithetal1976} attempted to rationalize their observations of shock instabilities in argon and carbon dioxide using the DK mechanism by quantitatively evaluating the DK instability criteria using both experimental measurements of the shock Hugoniot and equilibrium models.  They however observed shock instabilities well in the region predicted to be stable.  Shock instabilities have also been attributed to hydrodynamic instabilities in the low isentropic exponent gases in the presence of large shears, such as those due to large shock curvatures \cite{Hornung&Lemieux2001}. It is however difficult to reconcile this explanation with the instability observed in planar shock tube experiments.  In spite of these difficulties to account for the experimental observations, shock instabilities have been reproduced in numerical experiments for ionizing shocks in argon by direct numerical simulation \cite{Kapper&Cambier2011, Mondetal1997}.  These simulations attribute the instabilities to the coupling of the nonlinear kinetics of the collisional-radiative model with wave propagation within the induction zone, akin to the mechanism of cellular formation in gaseous detonations \citep{Fickett&Davis1979}.  Although the physical mechanism remains unclear, these numerical findings suggest that the observed instabilities in previous experiments are not due to an experimental artefact, such as the contact surface Taylor instabilities prevalent in experiments.\\
The present study attempts to elucidate the mechanism at play common to these previous experiments.  To formulate our hypothesis, we first note that the common feature of the unstable configurations summarized above are that the compressed medium experiences a dissipative process during the relaxation to the equilibrium post-shock state.  In strong shock waves, the first internal modes to equilibrate within the shock structure are the translational and rotational degrees of freedom \cite{Vincenti&Kruger1975}, within a few mean free paths.  If the shock wave is sufficiently strong, vibrational modes can also be excited, but these take a significantly larger number of mean free times to equilibrate \cite{Vincenti&Kruger1975}.  Likewise, if the temperature is sufficiently high to permit ionization and dissociation, the energy available in the translational and rotational modes is slowly patly transferred to these other internal modes through inelastic collisions.  All these energy relaxation processes correspond to an interval during which the translational and rotational kinetic energies of the molecules are lost through inelastic collisions.  It is this inelasticity in the collision process that we wish to model. Indeed, Goldhirsch and Zanetti have shown that a clustering instability is possible in initially uniform gases undergoing inelastic collisions \cite{Goldhirsch&Zanetti1993} .  In the present study, we wish to test whether the same instability mechanism also operates in dissipative gases undergoing shock wave relaxation to equilibrium. \\
In order to capture the key relaxation processes involved, and retain a sufficiently simple model which will permit us to gain further analytical insight, we assume a two-dimensional medium composed of hard disks.  We choose hard disks such that the simulations can be sufficiently simple to implement, test, analyze, visualize and reproduce. Furthermore, we have previously validated an analytical formulation of the shock jump conditions in a hard disk gas \cite{Sirmasetal2012}. This also permits us to analytically evaluate whether the DK and BZT mechanisms can account for the instability.\\
The model assumes that each binary collision is elastic, unless an \textit{activation} threshold is reached.  Activation is assumed to occur when the collision between two disks is sufficiently strong.  This mimicks the excitation of higher degrees of freedom (rotation, vibration, dissociation, ionization, etc...) with increasing temperatures \cite{Vincenti&Kruger1975}.  Quantitatively, the collision between two disks is assumed to be elastic if the relative speed between two disks taken along the line of action is below a threshold $u^*$, a classical activation formalism in chemical kinetics. For collisions with a higher amplitude, we assume an inelastic dissipative collision modelled with a constant coefficient of restitution $\epsilon<1$.  The problem we study is a classical shock propagation problem, whereby the motion of a suddenly accelerated piston driven in a thermalized medium drives a strong shock wave.  The piston is initially at rest and suddenly acquires a constant velocity $u_p$.  This model allows for the dissipation of the non-equilibrium energy accumulated within the shock structure which terminates once the collision amplitudes fall back below the activation threshold.  In this manner, the activation threshold also acts as a tunable parameter to control the equilibrium temperature in the post shock media, see below.  Note that the model assumed is also the standard model for granular gases \cite{Brilliantov&Poschel2004}, and our results may find applications in that field.\\
The numerical experiments thus reconstruct the dynamics of the hard disks.  These are calculated using the Event Driven Molecular Dynamics technique first introduced by Alder and Wainright \cite{Alder&Wainright1959}.  We use the implementation of Poschel and Schwager \cite{Poschel&Schwager2005}, that we extended to treat moving walls.  The code was tested for non-dissipative media in our previous study \cite{Sirmasetal2012}.  The numerical experiments were performed using 30,000 disks, unless otherwise noted.  The particles were initialized with equal speed and random directions.  The system was let to thermalize and attain Maxwell-Boltzmann statistics.  Once thermalized, the piston started moving with constant speed.  Below, all distances have been normalized by the radius of the disks and speeds by the initial speed of the disks. The initial packing factor of the disks was chosen to be $\eta_1=(V_a/m)/v=0.012$, where $V_a/m$ is the specific volume (area) of the hard disk and $v$ the specific volume; the initial gas is thus in the ideal gas regime \cite{Sirmasetal2012}.\\
The model was found to predict shock instabilities.  Figure \ref{fig:stream} shows the structure of the compressed gas and its macroscopic averaged properties for a piston propagating with $u_p=7$, with $\epsilon=0.95$ and an activation threshold of $u^*=3$.   For reference, the shock Mach number is approximately 6.6.   As can be clearly seen, the shock acquires a bumpy structure of approximately $10-15 \lambda_1$ spacing, where  $\lambda_1$ is the mean-free-path evaluated at the initial state. Animation of the dynamics of the particles showed the evidence of macroscopic convection rolls.  These were confirmed by coarse-graining the particle dynamics in order to visualize the streamlines.  As can be seen in Figure \ref{fig:stream}, the large bumpy structure corresponds to vortex rolls, where the material accumulates in the large scale high density clusters.  We have performed calculations with different domain sizes and number of particles, but this did not change the manifestation of the instability nor its characteristic spacing.\\
Figure \ref{fig:stream}b shows the coarse-grained density, temperature and pressure within the shock layer, obtained at the same evolution time as figure \ref{fig:stream}a.  These were obtained by ensemble averaging and coarse-graining in strips of width $0.5 \lambda_1$.  Note that density and temperature were obtained directly from the particle data, while pressure was estimated using the Helfand equation of state for a dense hard disk medium \cite{Helfand&Frisch1961}. \\
The one-dimensional average shock structure resembles the characteristic structure of relaxing media: the translational temperature has a peak within the shock and then decays while the non-equilibrium shock state dissipates its energy.  For reference, Figure \ref{fig:stream}b also shows the temperature evolution in a shock propagating through an elastic medium at the same initial conditions and shock Mach number (i.e., same mass flux across the shock).  As can be seen, the peak in the dissipative calculation corresponds closely to the equilibrium translational temperature.  Our model thus recovers the classical picture of relaxation phenomena inside shock waves (see Ref.  \cite{Zeldovich&Raizer1966} for example), where the first modes to relax are the translational ones. The subsequent drop in temperature in the dissipative calculation is due to the inelastic collisions, which eventually become elastic again once the majority of the collisions become de-activated.\\
Further insight into the shock transition structure and instability  mechanism can be deduced by constructing the equilibrium shock Hugoniot, that is the locus in the pressure-volume plane of possible end states obtained after shock compression.  Figure \ref{fig:hugoniot} shows our numerical results obtained for different shock strengths for elastic collisions ($\epsilon=1$) and inelastic collisions ($\epsilon=0.95$).  Also shown are the Hugoniot curves obtained analytically, using the Helfand equation of state for a hard disk medium \cite{Sirmasetal2012}. For the dissipative medium, the Hugoniot prediction captures very well the numerical results, as expected by our selection of the activation threshold to ensure an approximately isothermal system.  Also shown in Figure \ref{fig:hugoniot} is the evolution of the medium's state from initial to equilibrium state, for conditions illustrated in Figure \ref{fig:stream}; this transition line is termed the Rayleigh line.\\  
The shock Hugoniot and Rayleigh lines configurations obtained can immediately rule out any BZT behavior \cite{Zeldovich&Raizer1966}.  This would require a convex Hugoniot curve in the vicinity of the transition from elastic to inelastic behavior.  The Hugoniot remains concave.  Furthermore, BZT behavior would require the formation of a double discontinuity \cite{Zeldovich&Raizer1966}, which again is not the case; the transition from initial to equilibrium state is achieved across a single discontinuity, since the Rayleigh line does not exhibit any convex behaviour.  Since the slope of the line gives the mass flow rate across the shock wave, the constancy of the mass flow rate signifies a single discontinuity with a unique speed.  Note however the slight concavity of the Rayleigh line, a signature of viscous stresses in the shock wave \cite{Zeldovich&Raizer1966}.\\
We can also determine whether DK instability is expected or not based on the equilibrium shock Hugoniot data. The necessary condition for DK instability requires that the D'yakov parameter, $h$, be greater than a critical value $h_c$, given by 
\begin{align}
h=j^2 \left(\frac{dV}{dp} \right)_H > h_c=\frac{1-M_2^2(1+V_1/V_2)}{1-M_2^2(1-V_1/V_2)} \label{eq:DKcrit}
\end{align} 
where the derivative is taken along the equilibrium shock Hugoniot and estimated at the equilibrium state $2$, while $j^2=(p_2-p_1)/(V_1-V_2)$ is the square of the mass flux density across the shock.  For the dissipative hard disk gas considered, an isothermal equilibrium state approximates very well the Hugoniot.  Using the isotherm to evaluate the derivative, we find that the DK criterion is never satisfied.  The results are shown in Figure \ref{fig:DKcheck}.  Neither the elastic nor the dissipative medium is expected to develop DK instabilities.  We can thus conclude that neither BZT behavior nor the DK instability are compatible with the results of the numerical experiments.\\
To investigate whether the instability observed is compatible with the clustering instability of homogeneous dissipative gases \cite{Goldhirsch&Zanetti1993}, we conducted a separate simulation where all the gas is initialized with an average speed of $u=7$ and random directions.  This velocity was chosen to correspond to the piston speed investigated above.  Once initiated, we did not include an activation threshold, and let the system cool via the homogeneous cooling state of granular dynamics (see \cite{Brilliantov&Poschel2004}).  We found that large scale density clusters appear on time scales only a few times longer than observed for the piston driven shock flows described above.  While the time scales of the clustering formation in the homogeneous cooling state were indeed expected to be longer than in the piston driven case, since the piston continuously energizes the shocked gas, the similarity in the time scales suggest that the same clustering instability is at hand.  The large density clusters are formed by the inability of the gas within the cluster to leave owing to the local high rates of energy dissipation \citep{Goldhirsch&Zanetti1993}.\\
To gain further insight into the role of the rate of dissipation in the instability mechanism, we have further conducted numerical experiments by varying $\epsilon$. Varying $\epsilon$ does not change the equilibrium final state, and does not affect the DK predictions.  It only changed the rate at which energy is dissipated \cite{Brilliantov&Poschel2004}.  Figure \ref{fig:othereps} shows the shock structure obtained with different values of $\epsilon$.  We find that $\epsilon$ controls both the relaxation length and the spacing of the bumps, which, to first approximation, was found to be approximately given by the relaxation length scale.  This further supports our conclusion that the instability is associated with the rate of dissipation leading to cluster formation \cite{Goldhirsch&Zanetti1993}.\\
The numerical experiments conducted have thus shown that shock instabilities in relaxing media can be accounted for via the Goldhirsh and Zanetti clustering instability mechanism in dissipative gases, when the D'yakov-Kontorovich and Bethe-Zel'dovich-Thompson behavior can be ruled out.  However, a quantitative prediction of the experimental observations in relaxing gases requires a much more realistic relaxation model, which we leave for further study. \\
We gratefully acknowledge partial financial support from DRDC-Suffield (Dr. J. J. Lee as technical authority), NSERC through a Discovery Grant and the University of Ottawa through an Initiation of Research grant.  N.S. is a recipient of an Ontario Graduate Scholarship.

\bibliography{references}

\end{document}